\documentstyle [12pt]{article}
\begin{document}

\hfill {WM-96-105}

\hfill {May 1996}

\vskip 1in   \baselineskip 24pt
{

   \bigskip
   \centerline{\bf $K_L\rightarrow \pi^o\nu\overline{\nu}$ in Extended Higgs Models }
 \vskip .8in
 
\centerline{Carl E. Carlson, Greg D. Dorata and Marc
Sher } 
\bigskip
\centerline {\it Physics Department, College of William and
Mary, Williamsburg, VA 23187, USA}

\vskip 1in
 
{\narrower\narrower The decay $K_L\rightarrow \pi^o\nu\overline{\nu}$ is an 
excellent probe of the nature of CP
violation.  It is almost entirely CP-violating, and hadronic uncertainties are negligible. 
Experiments which hope to detect the decay are currently being planned. We calculate the
decay rate in several extensions of the standard model Higgs sector, including the
Liu-Wolfenstein two-doublet model of spontaneous CP-violation and the Weinberg
three-doublet model. In a model with an extra doublet, with CP-violation arising from the CKM
sector, the rate can increase by up to 50\%.  However, in models in which the CP
violation arises either entirely or predominantly  from the Higgs sector, we find that the
decay rate is much smaller than that of the standard model, unless parameters of the model are
fine-tuned.}

\newpage

\section{Introduction}

One of the deepest mysteries in theoretical physics concerns the nature 
and origin of the CP violation observed in the kaon system.   Although 
it can be accommodated within the three generation standard model, most 
extensions of the standard model contain additional sources of CP 
violation\cite{jarlskog}.  A primary motivation for the construction of 
B-factories is to explore CP violation in a regime in which it is 
expected to be considerably larger.

Within the standard model, much of the effort in understanding CP 
violation has focused on finding the values of the CKM matrix, which are 
parameterized by the Wolfenstein parameterization\cite{wolf,blg}, in which one has
\begin{equation}
|V_{us}|=\lambda;\quad |V_{cb}|=A\lambda^2;\quad 
V_{ub}=A\lambda^3(\rho-i\eta);
\quad V_{td}=A\lambda^3(1-\overline{\rho}-i\overline{\eta})
\end{equation}
where $\overline{\rho}\equiv \rho(1-\lambda^2/2)$ and 
$\overline{\eta}\equiv \eta(1-\lambda^2/2)$.  From K- and B- decays, 
these parameters can be determined.
Unfortunately, the interpretations of many current measurements of CP 
violation in the kaon system, as well as of many future measurements in 
the B system, are plagued by theoretical uncertainties.  These result 
from the absence of precise non-perturbative calculations of hadronic 
matrix elements.  For example, determination of $V_{cb}$ and  $V_{ub}$ 
to an accuracy of better than 5\% and 10\% respectively may not be 
possible without a significant improvement in the determination of 
hadronic matrix elements.  In addition, loop-induced decays also contain 
significant theoretical uncertainties, which affect the predictions (and 
interpretations) for $\epsilon'/\epsilon$,
$B^0-\overline{B}^0$ mixing, etc.   As emphasized by Buras {\it et 
al.}\cite{burasa}, even with optimistic assumptions about the theoretical 
and experiment errors, it will be difficult to achieve an accuracy 
better than $\pm 0.15$ in $\rho$ and $\pm 0.05$ in $\eta$.

As also emphasized by Buras and others\cite{burasb}, there are two processes 
in which the hadronic uncertainties are significantly reduced, and two 
processes in which they are virtually absent.  The former two are 
$K^+\rightarrow \pi^+\nu\overline{\nu}$ and the ratio of 
$B^0_d-\overline{B}_d^0$ mixing to $B^0_s-\overline{B}_s^0$ mixing.  The 
gold-plated decays, in which theoretical uncertainties are extremely 
small, are the CP asymmetry in $B^0_d \rightarrow \psi K_s$ and the 
decay $K_L\rightarrow \pi^o\nu\overline{\nu}$.   Buras\cite{burasa} has 
noted that measurement of the CP asymmetry in $B^0_d \rightarrow \psi K_s$ plus a 
measurement of the branching ratio for $K_L\rightarrow 
\pi^o\nu\overline{\nu}$ would allow a determination of all of the 
elements of the CKM matrix without any significant hadronic 
uncertainties, assuming that the CP violation is entirely in the CKM matrix.
  In this paper, we will be concentrating on the mode 
$K_L\rightarrow \pi^o\nu\overline{\nu}$, which (up to $O(\epsilon)$ 
corrections) is entirely CP-violating and free of substantial hadronic 
uncertainties.

The expected branching ratio for $K_L\rightarrow \pi^o\nu\overline{\nu}$ 
in the standard model is approximately $3\times 10^{-11}$.  This is many 
orders of magnitude smaller than the current upper bound of 
$5.8\times 10^{-5}$\cite{current}.  However, upcoming experiments are expected to 
improve the bound to $10^{-8}$, and preliminary studies at CEBAF and 
BNL\cite{private} 
 claim  that, not only could $K_L\rightarrow \pi^o\nu\overline{\nu}$ be 
detected, but  as many as 100 events could be seen.  Although these 
studie
s are only in a very preliminary stage, a 10\% measurement of the 
branching ratio does not appear to be impossible within the next decade.

As discussed above, virtually all extensions of the standard model 
contain additional sources of CP violation.  One might expect the 
branching ratio for $K_L\rightarrow \pi^o\nu\overline{\nu}$ to be 
different in these models.  Although the branching ratio has been 
calculated in the standard model, including QCD 
corrections\cite{burasc}, we know of no calculations of the branching 
ratio in models in which CP violation arises from a source other than
the CKM matrix.

Given the potential precision of a measurement of $K_L\rightarrow 
\pi^o\nu\overline{\nu}$, and the likelihood of additional sources of CP 
violation in extensions of the standard model, it is important to 
calculate the branching ratio in these extensions.  Even if it is some 
time before the necessary precision is reached, one should still look at 
the branching ratio in extensions of the standard model in the hope that 
some models might have a significantly higher rate--this might motivate 
``intermediate" experiments which might not reach the standard model 
rate.  For example, the electric dipole moment of the neutron is very 
small in the standard model, and is not in reach of experiments, but 
extensions of the model can have a much larger rate, and this has 
provided strong motivation for experiments which have lowered the bound 
substantially (ruling out several models in the process).

In this paper, we will calculate the rate for $K_L\rightarrow 
\pi^o\nu\overline{\nu}$ in models with an extended Higgs sector.  Such 
models are the simplest extensions of the standard model, and have 
additional sources of CP violation.  Models with additional gauge 
groups, such as left-right models, and supersymmetric models, are 
currently under investigation.

In the next section, we will review the standard model result
for $K_L\rightarrow \pi^o\nu\overline{\nu}$, and then consider
the simplest extension of the standard model, in which a single
Higgs doublet is added to the standard model, and yet all of the
CP violation still arises from the CKM matrix.  In Section 3, the
most general two-doublet model in which CP is violated
spontaneously will be considered, along with the Weinberg three-doublet model; in one
subsection, the effects of neutral Higgs bosons will be considered and in the next
subsection, the effects of charged Higgs bosons will be included.  Finally,  in
Section IV, we present our conclusions.

\section{The standard model and simplest extension}

The calculation of $K_L\rightarrow \pi^o\nu\overline{\nu}$ amounts
to determining the coefficient of the effective Lagrangian for
$d\overline{s}
\rightarrow \nu\overline{\nu}$, and evaluating the hadronic matrix element.
The matrix element will be the same as that in semileptonic $K_L$ decay,
and thus in the ratio of the rate for $K_L\rightarrow \pi^o\nu\overline{\nu}$ to that of
the semileptonic decay, the matrix element will cancel.  There are two types of diagrams which
contribute to this effective Lagrangian.  The first are Z-penguins, generated by an induced
$d\overline{s}Z$ coupling, and are shown in Figure 1. The second consist of box diagrams, shown
in Figure 2.

Inami and Lim\cite{il} have calculated these contributions, in the limit that
external masses and momenta are much smaller than the internal masses.
The amplitudes are then described by an effective four-fermion interaction:
The effective Lagrangian is given by the form
\begin{equation}
{\cal{L}}_{eff}=-{G_F\over\sqrt{2}}{\alpha\over 4\pi\sin^2\theta_W} V^*_{ts}V_{td}
\left(
4\ D\overline{s}_L\gamma_{\mu}d_L\sum_{i=1}^3 \nu_{L_i}\gamma^{\mu}\nu_{L_i}
\right)
\end{equation}
where the sum is over the three neutrino flavors.
  Using unitarity,
Inami and Lim show that one can calculate the contribution due to the top
quark, and then (ignoring the up and charm quark masses) subtract  the
mass independent part, so that only the CKM matrix elements involving the top quark
enter.
In the CP-violating decay, $K_L\rightarrow\pi^o\nu\overline{\nu}$, the
imaginary part of ${\cal L}_{eff}$ will enter.

Including the contribution of the box diagrams (in the limit that lepton masses
are ignored compared to the top),
\begin{equation}
D(x_t) = {x_t\over 4}\bigg[{3x_t-6\over (1-x_t)^2}\ln\ x_t + {x_t+2\over x_t-1}\bigg]
\end{equation}
where $x_t\equiv m_t^2/M_W^2$.
The ratio of branching ratios is then
\begin{equation}
{B(K_L\rightarrow \pi^o\nu\overline{\nu})\over B(K^+\rightarrow \pi^oe^+\nu)}
= {3\over 4\pi^4}{\tau_{K_L}\over \tau_{K^+}}\left(G_F^2m_W^4\right)^2
A^4\lambda^8\eta^2D(x_t)^2
\end{equation}
which gives a branching ratio of $\sim 3\times 10^{-11}$ in the standard model (using
$\eta-0.35\
$(see ref.
\cite{burasa}).

We begin our consideration of extensions beyond the standard model by looking
at the simplest extension:  the two-doublet model, in which CP violation
occurs through the CKM sector.  In this case, the rate will also depend on 
the imaginary part of $V_{ts}^*V_{td}$, and the only change will be the addition
of charged Higgs loops in Figure 1.

In this simplest extension, one Higgs doublet couples to
one quark charge, and the other couples to the other
quark charge.  The detailed vertices and Lagrangian are
well-known\cite{hhg} (and can be obtained from the $\xi=0$ limit of the model
discussed in the next section).  The neutral
Higgs boson interactions are flavor-conserving, and thus will not
contribute to the diagrams of Fig. 1.  The only
difference is that we now have physical charged Higgs
bosons in the loop instead of just W and Goldstone bosons.
The charged Higgs bosons appear in diagrams (a), (b),
(d) and (h) (note that there is no ZWH vertex in the
model).  The divergences in these diagrams cancel, and we
find that the ratio of the contribution of charged Higgs
boson loops to the amplitude relative to the standard model
result, $R$ is \begin{equation} R=
-{1\over 4}\cot^2\beta{(1-x_t)^2\over(1-x)^2}\left(
{
(x(4-x)-2x^2\cos 2\theta_W)\ln\ x\ +\ x(1-x)(3-2\cos 2\theta_W)
\over
(3x_t-6)\ln\ x_t\ -\ (1-x_t)(2+x_t)}\right)
\end{equation}
where $x\equiv (m_t/m_{H^+})^2$ and 
 $\tan\beta$ is the ratio of the vacuum
expectation values of the two Higgs doublets (in most unified models it is
 greater than unity, and must be greater than
$0.5$ for perturbation theory to be valid).  For a charged Higgs mass of $150\
(250,400)$ GeV, the ratio is $R=.32\ (.20,.12)$ times $\cot^2\beta$.  Thus, for $\tan\beta$
near unity, this can increase the branching ratio by a factor of $1.74$ for a charged Higgs
mass of $150$ GeV.  It should be noted that this model has a lower bound on the charged Higgs
mass arising from $b\rightarrow s\gamma$ of $200$  GeV\cite{grant}, which gives an increase in
the branching ratio of approximately $50\%$ (for $\tan\beta\sim 1$).

Belanger et al.\cite{belanger}  have also considered the rate for $K_L
\rightarrow \pi^o\nu\overline{\nu}$ in this model .  Their results
are consistent with ours.  They note that the ratio of the rates
is given by $(1+R)^2Q$, where $Q$ is the ratio of the CKM parameters $(A^4\eta^2)$
as determined  from
experiments {\it including} the effects of the charged Higgs to the values of
these parameters as determined from experiments in the standard model (without the charged
Higgs).   The value of $Q$ is consistent with unity, since no discrepancy with the
standard model is seen.  However, by scanning parameter-space, and
requiring all experimental results to be within the 90\% confidence
level, they show that there is a region of parameter-space in which the
value of
$Q$ can be somewhat larger,  leading to a larger rate.  Our philosophy is that this
involves charged Higgs effects in  experiments other than $K_L
\rightarrow \pi^o\nu\overline{\nu}$, and that by the time the experiment
is done, the uncertainties in $(A^4\eta^2)$ will be much smaller, in the
range of 10 percent\cite{burasa}.  Nonetheless, one should be aware that
the extraction of the CKM angles in this model may give results different
from those in the standard model.

\section{Spontaneous CP-violation}

Another attractive mechanism for CP-violation is
spontaneous CP-violation\cite{lee}.  This cannot occur in
the single Higgs model, and thus requires extension of
the Higgs sector.  If one adds one more Higgs doublet,
then one can violate CP spontaneously, but at the cost of
tree level flavor-changing neutral currents(FCNC).  The
discrete symmetry that is usually implemented to
eliminate such currents will also eliminate the
spontaneous CP violation\cite{branco}.  One has two
choices:  break the discrete symmetry by parameters which
are sufficiently small that FCNC are not
phenomenologically problematic, or keep the discrete
symmetry and enlarge the Higgs sector by adding the third
doublet.  The former option was analyzed in detail by Liu
and Wolfenstein\cite{lw}, the latter is the model of
Weinberg\cite{weinberg}.  We first consider the Liu-Wolfenstein model.

The model contains two Higgs doublets, and the most
general CP-invariant Yukawa coupling and Higgs potential
is
\begin{eqnarray}
-{\cal{L}}_Y&=&\overline{\Psi}_{Li}^o(F_{ij}\tilde{\Phi}_2+
\xi F'_{ij}\tilde{\Phi}_1)U^o_{Rj}+
\overline{\Psi}^o_{Li}(G_{ij}\Phi_1+\xi
G'_{ij}\Phi_2)D^o_{Rj}+{\rm h.c.},\cr\cr
V&=&-\mu^2_1\Phi_1^\dagger\Phi_1-\mu^2_2\Phi_2^\dagger\Phi_2
+\lambda_1(\Phi_1^\dagger\Phi_1)^2+\lambda_2(\Phi_2^\dagger
\Phi_2)^2\cr
&+& \lambda_3(\Phi^\dagger_1\Phi_1)(\Phi^\dagger_2\Phi_2)
+\lambda_4(\Phi^\dagger_1\Phi_2)(\Phi^\dagger_2\Phi_1)
+{1\over 2}\lambda_5\left[(\Phi^\dagger_1\Phi_2)^2
+(\Phi^\dagger_2\Phi_1)^2\right]\cr
&+&{1\over
2}\xi'(\Phi^\dagger_1\Phi_2+\Phi^\dagger_2\Phi_1)(\lambda_6
\Phi^\dagger_1\Phi_1+\lambda_7\Phi^\dagger_2\Phi_2)
\end{eqnarray}

Here, $\xi$ and $\xi'$ are small parameters which
determine the amount by which the discrete symmetry
($\Phi_2\leftrightarrow -\Phi_2,\ D^o_R\leftrightarrow 
-D^o_R$) which eliminates FCNC is broken. The fact that
both Higgs doublets couple to all of the fermions ensures
the existence of FCNC, since diagonalizing the quark mass
matrix will not automatically diagonalize the Yukawa
coupling matrices.   Minimizing the potential yields
\begin{equation} \langle\Phi_1\rangle=\sqrt{1\over
2}\left({0\atop v_1}\right),\qquad
\langle\Phi_2\rangle=\sqrt{1\over 2}\left( {0\atop
v_2e^{i\alpha}}\right). \end{equation} The CP-violating
phase $\alpha$ is given by \begin{equation} \cos\
\alpha=-\xi'{\lambda_6v_1^2+\lambda_7v^2_2\over
4\lambda_5v_1v_2}. \end{equation} 

Liu and Wolfenstein discuss two limiting cases.  If
$\xi=0, \xi'\neq 0$, then the model becomes an earlier
model of Branco and Rebelo\cite{rebelo}.    Here,
CP-violation occurs in the Higgs sector, however, there
are no FCNC at tree level, and thus in order to obtain a
$\Delta S=2$ CP-violation one must go to two loops. 
As a result, the value of $\epsilon$ is too small.
The second case is if $\xi'=0, \xi\neq 0$, then the
CP-violating phase is $\pi/2$.  As Liu and Wolfenstein discuss, spontaneous
CP violation in this limit is the same as introducing a
purely imaginary Yukawa coupling $i\xi$ which breaks the
discrete symmetry. Although this model is certainly
viable, there is no natural mechanism for ensuring
$\xi'=0$, although they use this limit in their numerical
examples, as will we.

In this model, there will be  contributions to
the $K_L\rightarrow \pi^o\nu\overline{\nu}$ rate 
from charged Higgs loops (as in the simple model in the
last section, albeit with very different couplings), as
well as from neutral Higgs loops.  Since CP violation has
a different origin in this model, one might hope to avoid
the $V_{ts}^*V_{td}$ suppression factor present in the
standard model result.  

\subsection{Neutral Higgs bosons}

We will first consider
effects of neutral Higgs bosons.  Since the neutrinos are
very light, their interactions with Higgs bosons will be
negligible, and thus box diagrams will not contribute.  We
have only corrections to the $\overline{s}dZ$ vertex, and
the internal fermion line will be a $b$-quark, rather
than a top quark.   It is clear that we will need two
flavor-changing neutral current couplings, so the result
will be proportional to $\xi^2$.   

The flavor-changing Yukawa couplings can be found from the
Yukawa terms in Eq. 6.   The couplings of the neutral
complex fields, $\phi_1$ and $\phi_2$, to down-type quarks are given by
\begin{equation}
-{\cal
L}_Y=\overline{D'}_{Li}(G_{ij}\phi_1+\xi
G'_{ij}\phi_2)D'_{Rj}+{\rm h.c.}\end{equation}
where the primes indicate the weak eigenstate basis.
Plugging in $v_1/\sqrt{2}$ and $v_2e^{i\alpha}/\sqrt{2}$ for
the vacuum expectation values, and defining $D'_R\rightarrow
e^{-i\alpha} D'_R$ yields the mass matrix
\begin{equation}
M_d={1\over\sqrt{2}}(G+e^{-i\alpha}\xi{v_2\over
v_1}G')v_1\equiv M_d^o+e^{-i\alpha}\xi{v_2\over v_1}M_d'
\end{equation}
where flavor indices have been suppressed.
There are three neutral physical Higgs fields and one neutral Goldstone boson, which
we denote by
$H_j$, with
$j=1-4$ where $H_4$ is the Goldstone boson (the
calculation is done in the Feynman gauge, so the Goldstone
boson mass is the Z-boson mass).   To rotate to the fermion
mass eigenstate basis, we need to define
\begin{equation}
N\equiv V_LM'_{d}V_R^\dagger\end{equation}
where $V_{L,R}$ rotate $D'_L$ and $D'_R$ into their mass
eigenstates $D_L$ and $D_R$.
We then find that the general flavor-changing Yukawa coupling of
$\overline{D}_{Li}D_{Rj}H_k$ is given by
\begin{equation}
i{(\sqrt{2}G_F)^{1\over 2}\over \cos^2\beta}
\xi\overline{D_i}\left[ e^{i\alpha}N_{ij}(S_{2k}+iS_{4k})R
+e^{-i\alpha}N_{ji}^*(S_{2k}-iS_{4k})L\right]D_jH_k
\end{equation}
Here, $L$ and $R$ are ${1\over 2}(1\mp\gamma_5)$,
$\tan\beta\equiv v_2/v_1$ and $S_{ij}$ is the matrix which
diagonalizes the $4\times 4$ Higgs mass matrix.  $S_{ij}$
depends on parameters in the Higgs potential and is
essentially undetermined. Note that if $\xi'=0$, then the
$4\times 4$ matrix divides into two $2\times 2$ matrices (the
scalar and pseudoscalar matrices, respectively), and then
either $S_{2k}$ or $S_{4k}$ will vanish, greatly simplifying
the vertex.

Due to the proliferation of parameters, we will
 now greatly simplify the calculation by taking
the special case $\xi'=0$, as was done by Liu and
Wolfenstein.  There is a potential delicacy with that
limit.  If the Lagrangian is CP-invariant (i.e. all of
the CP-violation arises spontaneously), then only $\xi$
and $\xi'$ can violate CP.  Any effect proportional to
$\xi^2$ only will then not violate CP (as discussed
above, $\xi'=0$ is equivalent to multiplying $\xi$ by $i$).
However, one can certainly have a model in which there
is both explicit {\it and} spontaneous CP violation,
thus the $N$ matrices need not be real. In that case, our
results will not be significantly affected by this
assumption. Even if one assumes that the Lagrangian is
CP-invariant, and relaxes the $\xi'=0$ assumption, then there will be terms of
$O(\xi^2\xi')$, as well as $O(\xi^3)$, which do violate CP; these terms will be
$O(\xi')$ or $O(\xi)$ times terms that we will calculate.  In that particular case,
under the
assumption that the Lagrangian is CP-invariant, our numerical
results would be somewhat larger than the actual result
(note that there are no real bounds on the size of $\xi'$
other than it is ``small").  We will discuss the
implications of $\xi'\neq 0$ later.

The neutral Higgs loops contribute to diagrams (a), (b) and
(d) in Fig. 1., in which the $G^-$ is replaced by a neutral Higgs (and the
$u_i$ fermion is replaced by a $d_i$; the leading contribution will come from internal
$b$-quarks.  Note that under the assumption
$\xi'=0$, the scalars and pseudoscalars decouple, and the Z boson only
couples to a scalar plus a pseudoscalar.  As a result,
diagram (h) doesn't contribute to the vector
$\overline{s}\gamma_\mu d$ effective Lagrangian. In addition, the need for two
flavor-changing neutral current vertices implies that both fermion vertices must
involve a Higgs boson, and thus diagrams (f) and (g) will not contribute. A further
simplification, for the sake of illustration, can be made by taking all of the neutral scalars
to have the same mass as the Goldstone boson, i.e. $M_Z$--we will discuss the results of
relaxing this assumption shortly. In that case, the resulting sum over the four Higgs boson
contributions just becomes $\sum_k (S_{2k}^2+S_{4k}^2)$, which is $2$.  The effective
Lagrangian from these loops is found to be \begin{equation}
{\cal L}={G_F\over
4\sqrt{2}\cos^4\beta}{\alpha \over 4\pi\sin^2\theta_w}T_1 
\overline{s}\gamma_\mu d \overline{\nu}\gamma^\mu \nu
\end{equation}
where 
\begin{equation}
T_1=\xi^2
{N_{sb}N^*_{db}-N^*_{bs}N_{bd}\over m^2_W}\left(
{x_b(4-x_b)\ln(x_b)\over (1-x_b)^2}+{3x_b\over 1-x_b}\right)\end{equation}
and $x_b\equiv m_b^2/m^2_H$.  Note that if we relax the
assumption that the Higgs masses will be the Z mass, but
still assume that they are degenerate, then this result will
hold except for a slightly different contribution from the
Goldstone boson.  One expects, of course, the lightest of
the Higgs bosons to give the biggest contribution. 

Using this result, we can find the ratio of amplitudes, ${\cal A}$ for 
$K_L\rightarrow \pi^o\nu\overline{\nu}$ in this model to that in the standard
model. This gives

\begin{equation}
{{\cal A}_{new}\over {\cal A}_{SM}}=
{\xi^2\over\cos^4\beta}{{\rm Im}(N_{sb}N^*_{db}-N^*_{bs}N_{bd})
\over m^2_b}\quad T_2
\end{equation}
where
\begin{equation}
T_2\equiv {m^2_b\over 2m^2_W}{1\over {\rm Im} (V^*_{ts}V_{td})}
{x_b\over x_t}{[{(4-x_b)\ln(x_b)\over (1-x_b)^2}+{3\over (1-x_b)}
]\over [ {(3x_t-6)\over (1-x_t)^2}-{(2+x_t)\over (1-x_t)}]}
\end{equation}
Using the Wolfenstein parametrization, the ${\rm Im}V^*_{ts}V_{td}
$ term is $A\eta\lambda^5$, which is (for $\eta\simeq .35$) $1.8\times
10^{-4}$.  Using neutral Higgs masses of $m_Z$, as discussed
earlier, we find
\begin{equation}
{{\cal A}_{new}\over {\cal A}_{SM}}=
.06\ {\xi^2\over\cos^4\beta}{{\rm Im}(N_{sb}N^*_{db}-N^*_{bs}N_{bd}
)\over m^2_b}\end{equation}
At first sight, it appears that this ratio could be quite large. In
virtually all models, the value of $\tan\beta$ ranges from unity to
$m_t/m_b\sim 35$.  At the upper end of the range, $\cos^4\beta$ can
be as small as $10^{-6}$.  If $\xi\sim 0.1$, and the $N$ matrix elements
are the size of the largest mass scale expected ($m_b$), then the
ratio could be several hundred, leading to a rate as much as five orders
of magnitude greater than the standard model rate.

However, the value of $\xi N$ is not arbitrary.  It contributes to $\epsilon$
and thus is constrained. Liu and Wolfenstein have calculated the neutral Higgs contribution to
$\epsilon$.  In the two-generation case, they find,  taking the Higgs scalar masses to be
100 GeV,
\begin{equation}
{\xi^2\over\cos^4\beta}={2\times 10^{-3}\over\cos^{2/3}\beta\sin^{2/3}\beta}
\left({1\over(\sigma+\sigma')^{2/3}}{m_dm_s\over(N_{12}-N_{21})^{4/3}(N_{12}+N_{21})^{2/3}}\right)
\end{equation}
where one writes $N_{ij}$ in terms of its real and imaginary parts: 
$N_{ij}=N'_{ij}+i\xi\tan\beta\ n_{ij}$ and defines
\begin{equation}
\sigma\equiv-{n_{12}+n_{21}\over N'_{12}+N'_{21}}\qquad\sigma'\equiv{n_{21}-n_{12}\over 
N'_{12}+N'_{21}}.
\end{equation}
Since physical quantities can only depend on the  product $\xi N$, the expressions for
$\sigma$ and $\sigma'$ depend on a particular convention.  Liu and Wolfenstein scale $\xi$ by
assuming that
$N_{12}-N_{21}= m_s\sin\theta_c\simeq \sqrt{m_dm_s}$. With this convention, they  argue
that the natural values of
$\sigma$ and
$\sigma'$ are of
$O(1)$, and that if one assumes that the $N$ matrices have the same structure as the quark mass
matrices, then all of the terms in parentheses in Eq. (18) should be of $O(1)$.  Writing the
terms in parentheses ar $A'$, we can then write (with this convention)

\begin{equation}
{{\cal A}_{new}\over {\cal A}_{SM}}= 1.2\times 10^{-4}\ A' {1\over \cos^{2/3}\beta
\sin^{2/3}\beta}{{\rm Im}(N_{sb}N^*_{db}-N^*_{bs}N_{bd})\over m^2_b}
\end{equation}
Of course, the expression in Eq. (18) is only valid in the two-generation case.  In the
general case, the expression in parentheses will be much more complicated.  Nonetheless, the
result in Eq. (20) will be unaltered, and one still also expect the value of $A'$ to be $O(1)$.

Even if $\tan\beta\sim m_t/m_b$, this ratio will be no greater than one percent,
and thus unmeasurable.  The only way to get a large rate would be to assume that
either $A'$ is much greater than unity (which requires extensive fine-tuning)
or that the off-diagonal terms in the $N$ matrix are much larger than the
largest mass scale in the down-quark sector.
Neither of these seems likely.  In addition, the requirement that Higgs
mediated $B-\overline{B}$ mixing not be too large gives strong constraints\cite
{yao}
on $N_{bd}$, which we find to be approximately $\xi N_{bd}/m_b \leq 0.007$,
which further constrains the ratio.

\subsection{Charged Higgs Bosons}

What about the contribution of charged Higgs bosons in the Liu-Wolfenstein model?  In
this model, the coupling of the charged Higgs bosons to fermions is given by
\begin{equation}
{\cal L}=-i(2\sqrt{2}G_F)^{1/2}\left(
H^+\overline{U}(\Gamma_1L+\Gamma_2R)D+H^-\overline{D}(\Gamma_1^\dagger R
+\Gamma_2^\dagger L)U\right)
\end{equation}
where
\begin{eqnarray}
\Gamma_1&=& V_L^\dagger[\cot\beta M_u - \xi e^{i\alpha}N_u/\sin^2
\beta]\cr\cr
\Gamma_2&=& [\tan\beta M_d -
 \xi e^{-i\alpha}N_d/\cos^2
\beta]V_L^\dagger
\end{eqnarray}
Here, the matrices $M_u$ and $M_d$ are diagonal, and the matrices $N_u$
and $N_d$ are defined as in the neutral Higgs case.  $V_L$ is the CKM matrix.
Note that if $\xi=0$, the couplings reduce to the usual two-Higgs model.

The diagrams are the same as in the two-Higgs case, and only internal top
quarks are considered.  The effective Lagrangian arising from diagrams (a), (b) and
(d) is found to be
\begin{equation}{\cal L}_1=
{G_F\over
8\sqrt{2}}{\alpha \over 4\pi\sin^2\theta_W} T_1
\overline{s}\gamma_\mu d \overline{\nu}_L\gamma^\mu \nu_L
\end{equation}
where
\begin{equation}
T_1={({\Gamma_1})_{st}^\dagger(\Gamma_1)_{td}-
({\Gamma_2})_{st}^\dagger(\Gamma_2)_{td}
\over m^2_W}\left(
{x(4-x)\ln\ x\over (1-x)^2}+{3x\over 1-x}\right)\end{equation}
and  $x\equiv m^2_t/m^2_{H^+}$.  In this case, diagram (h) also contributes, and
the effective Lagrangian is 
\begin{equation}{\cal L}_2=
{G_F\over
4\sqrt{2}}{\alpha \over 4\pi\sin^2\theta_W} T_2
\overline{s}\gamma_\mu d \overline{\nu}_L\gamma^\mu \nu_L
\end{equation}
and 
\begin{equation}
T_2=-\cos 2\theta_W{({\Gamma_1})_{st}^\dagger(\Gamma_1)_{td}+
({\Gamma_2})_{st}^\dagger(\Gamma_2)_{td}
\over m^2_W}\left(
{x^2\ln\ x\over (1-x)^2}+{x\over 1-x}\right)\end{equation}

The  $\xi=0$ part of the effective Lagrangian is identical to the simplest extension
considered in the last section, in which there is no spontaneous CP-violation and
the CKM matrix is real.  What about the 
$\xi^2$ terms?  The
$\Gamma$ factors become
\begin{equation}
({\Gamma_1})_{st}^\dagger(\Gamma_1)_{td}\pm
({\Gamma_2})_{st}^\dagger(\Gamma_2)_{td}
=\xi^2\left({(V_LN_u^\dagger)_{st}(N_uV_L^\dagger)_{td}\over \sin^4\beta}
\pm{(N_d^\dagger V_L)_{st}(V_L^\dagger N_d)_{td}\over \cos^4\beta}\right)
\end{equation}
Once again, we don't know the values of the $N_u$ and $N_d$ matrix elements, but can
assume that they are not much larger than the top and bottom masses, respectively.
Consider the contribution of the $N_d$ terms.  They give an expression which is
identical to that of the neutral case except for some extra $V_L$ matrices and
replacing $x_b=m_b^2/m_H^2$ with $x=m^2_t/m^2_{H^+}$.  This latter change will reduce
the size of the final result (due to the absence of the large logarithm), and it is
unlikely that including the CKM matrices will increase the result, and thus the
contribution of the $N_d$ terms will also be very small.   The ratio of the
contribution of the $N_u$ terms to the standard model result is (choosing $m_{H^+}$
= 150 GeV and using Eq. (18))
\begin{equation}
\bigg|{{\cal A}_{new}\over {\cal A}_{SM}}\bigg| \simeq  10^{-5} A'
{\cos^{10/3}\beta\over
\sin^{14/3}\beta} {Im [(V_LN_U^\dagger)_{st}(N_uV_L^\dagger)_{td}]\over m^2_b}
\end{equation}
Even if one chose to ignore the CKM factors, and assume that $N_u$ is of order $m_t$,
then, since $\tan\beta\geq 1$, this is no more than $0.02 A'$, and thus will also
not be large (unless, as discussed earlier, one fine-tunes to make $A'$ large.
We conclude that the $\xi^2$ effects are not significant.

There is a cross-term which is linearly dependent on $\xi$  We find that 
\begin{equation}
\bigg|{{\cal A}_{new}\over {\cal A}_{SM}}\bigg| \simeq  10^{-2}\ \sqrt{A'}
{\cos^{8/3}\beta\over
\sin^{10/3}\beta} {m_t [V_{st} (N_uV_L)_{td} +V_{dt}(N_uV_L)_{ts}]\over m_b^2}
\end{equation}  Again,  if one assumes that $(N_uV_L)_{td}$ is approximately
$m_tV_{td}$, this is approximately $3\times 10^{-3} \sqrt{A'}$, which is not
measurable\cite{foo}

If one assumes that the CP violation is entirely spontaneous, i.e. that 
there is no CKM CP-violation, then this model has the ability, as shown by
Liu and Wolfenstein, to explain all observed CP-violating phenomena.  However,
as we have seen, it will generally give a much smaller rate for $K_L
\rightarrow \pi^o\nu\overline{\nu}$ than the standard model.  Note that, as discussed
earlier, if one does not assume $\xi'=0$, then the result will be $O(\xi)$ or
$O(\xi')$ times smaller than the terms that we have calculated.

Perhaps the most well-known model of spontaneous CP violation is the Weinberg
model\cite{weinberg}.  Although bounds from the neutron electric dipole moment and
$b\rightarrow s\gamma$ seem to rule out the model\cite{wein}, it might survive with some
fine-tuning and other similar models might still be viable.  This model assumes that there
are no tree-level flavor changing neutral currents, and as a result three Higgs doublets are
needed in order to violate CP spontaneously. All CP violation is to come from the Higgs sector,
and thus the CKM matrix is real.  Since there are no tree-level flavor changing neutral
currents, neutral Higgs bosons will not contribute to the
$K_L\rightarrow
\pi^o\nu\overline{\nu}$ decay at one-loop.  There are two charged Higgs bosons (in
addition to the charged Goldstone boson), whose couplings to fermions are given by
\begin{equation}
{\cal L}_Y=(2\sqrt{2}G_F)^{1/2}\sum_{i=1}^2 \left(
\alpha_i\overline{U}_LV_LM_DD_R+\beta_i\overline{U}_RM_UV_LD_L\right)H_i^++{\rm h.c.}
\end{equation}
where $V_L$ is the real CKM matrix.  The CP violation occurs in the (complex)
parameters $\alpha_i$ and $\beta_i$.  The observed CP violation parameter $\epsilon$
is proportional to $\sum_i\ {\rm Im}\ (\alpha_i\beta_i^*)/ m^2_{H_i^+}$.  Since the
neutron electric dipole moment is proportional to the same parameter, it is predicted in the
 model (modulo long-distance effects), and, as discussed above, tends to give too large a
value\cite{wein}.

In the calculation of the contribution of the charged Higgs bosons to the 
diagrams in figure 1, we find that all terms are proportional to $\alpha_i^*\alpha_i$
or to $\beta_i^*\beta_i$, and thus have no imaginary part; the one-loop penguin contributions
vanishes.  This is not surprising, since the value of $\epsilon$ and of the neutron
electric dipole moment involve the operator $\overline{d}\sigma_{\mu\nu}s$ whereas we
are here interested in $\overline{d}\gamma_\mu s$, and the extra $\gamma$ matrix is
needed to give the $\alpha_i\beta_i^*$ structure instead of $\alpha_i\alpha_i^*$.  There will
be a one-loop box contribution, but this will be suppressed by two powers of the tau-lepton
mass divided by $M_W$.   Thus the rate for
$K_L\rightarrow
\pi^o\nu\overline{\nu}$ in the Weinberg model will be much lower than that of the
standard model.

\section{Conclusions}

The process $K_L\rightarrow \pi^o\nu\overline{\nu}$ is an extremely promising probe
of the nature of CP violation.  It is almost entirely CP-violating and is free of
significant hadronic uncertainties.  The branching ratio, which is calculated quite
precisely in the standard model, is small, but within reach of currently planned
experiments, and its measurement to $10\%$ accuracy may be
possible.   In this paper, we have calculated the branching ratio in models in which
the CP violation arises either completely or partially from an extended Higgs sector.
We have concentrated on the Liu-Wolfenstein and Weinberg models, although the results
should be fairly general.  In spite of potentially large contributions, it has been
shown that when the constraints caused by fitting the value of $\epsilon$ are
included, the contribution of both neutral and charged Higgs bosons to the branching
ratio become very small.  Thus, in a model in which most or all of the CP violation
arises from the Higgs sector, the branching ratio for
$K_L\rightarrow \pi^o\nu\overline{\nu}$ will be much smaller than the standard
model result, and thus unmeasurable.  

We thank David Atwood for several useful discussions.  This work was supported by the
National Science Foundation grant No. NSF-PHY-9306141.

\def\prd#1#2#3{{\rm Phys.~Rev.~}{\bf D#1}, #2 (19#3)}
\def\plb#1#2#3{{\rm Phys.~Lett.~}{\bf B#1}, #2 (19#3)}
\def\npb#1#2#3{{\rm Nucl.~Phys.~}{\bf B#1}, #2 (19#3)}
\def\prl#1#2#3{{\rm Phys.~Rev.~Lett.~}{\bf #1}, #2 (19#3)}

\bibliographystyle{unsrt}

\begin{thebibliography}{99}
\bibitem{foo}
Of course, in the unlikely event that this corner element of the $N_uV_L$
matrix is extremely large (of the order of the largest element of the matrix), then the
effect would be measurable.
\end{thebibliography}

\newpage

\begin{figure}

\vglue 7.5in  
\hskip 1.00in {\special{picture penguins scaled 1000}} \hfil
\vglue 0.1in

\caption{Corrections to the $\overline{s}dZ$ vertex in the standard model.  $G$ refers to the
charged Goldstone boson.}
\end{figure}

\begin{figure}

\vglue 7.7in  
\hskip 1.25in {\special{picture boxes scaled 1000}} \hfil
\vglue 0.1in

\caption{Box diagrams contributing to $K_L\rightarrow\pi^O\nu\overline{\nu}$.}
\end{figure}

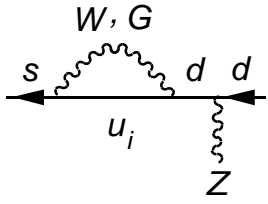
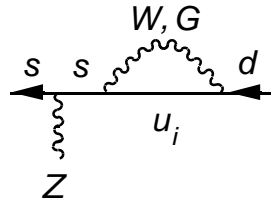
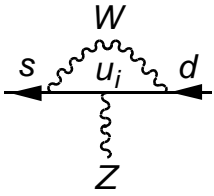
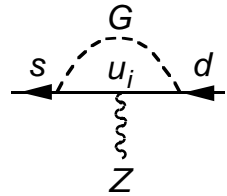
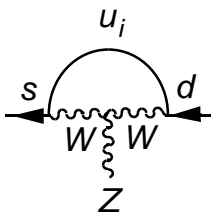
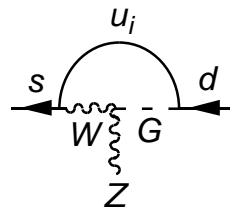
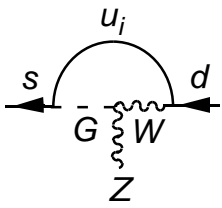
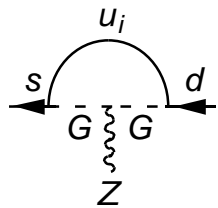
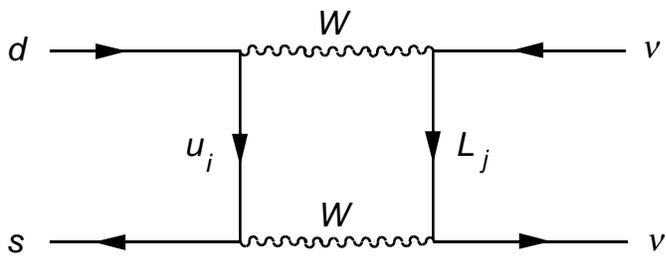
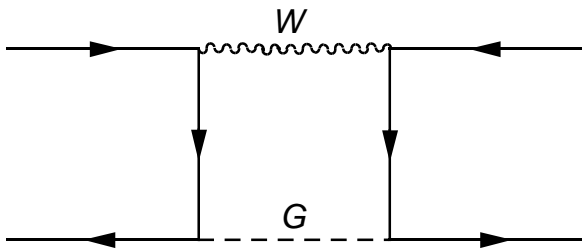
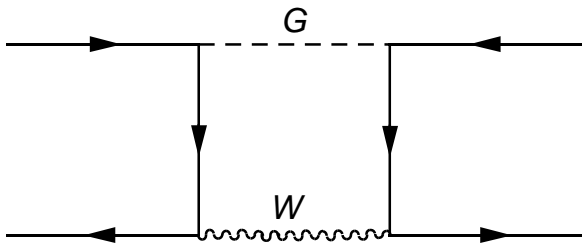
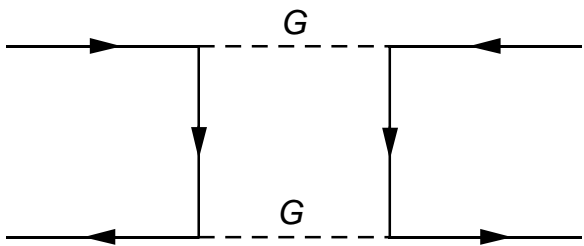
\end{document}